# МОДЕЛ НА ИНФОРМАЦИОННА СИСТЕМА ЗА УПРАВЛЕНИЕ НА ИНФОРМАЦИОННИ ПОТОЦИ

Радослава Кралева, Велин Кралев

Югозападен университет "Неофит Рилски", гр. Благоевград

*Резюме: В настоящата статия са разгледани някои проблеми при разработването на софтуерни продукти свързани с управлението на информационни потоци. Представени са основните етапи при тяхното разработване. Създадена е методика за концептуално моделиране и проектиране на информационни системи от такъв тип. С цел доказване ефективността на предложения модел е разработена информационна система за административното обслужване на докторанти във ВУЗ.*

**Ключови думи:** информационни потоци, софтуерен продукт, концептуален модел

## УВОД

Проектирането и реализирането на един софтуерен продукт може да бъде сравнено с проектирането и построяването на една сграда, тъй като се състои от множество неделими елементи. Програмния продукти все по-често заемат основна роля в дейността на дадена институция, било то финансова, образователна, търговска и т.н. В практиката този вид продукти се наричат корпоративни приложения [1], или информационни системи. Използваните информационни потоци се представят по различен начин, а потребителския интерфейс е необходимо да бъде дружелюбен и интуитивен. Същността на корпоративните приложения е управляването на големи масиви от данни. Разработването, съхраняването и управлението на релационни бази от данни е втората им отличителна черта.

В повечето български университети вече са създадени и внедрени различни програмни продукти, които да улеснят и автоматизират цялостната функционална дейност на административния персонал, на академичния състав и на самите обучаващи се. Основната цел, на която се подчинява развитието на информационното обслужване и инфраструктурата във висшето образование, е да се подобри качеството на обучението и административното обезпечаване на основните дейности извършвани в университетите. Съгласно чл. 17, ал. 2, т. 9 от ЗВО всеки университет трябва да разполага с университетски информационен център за обслужване на студенти и докторанти [2].

С настоящата статия се цели да се представи модел за реализация на информационна система, която да улесни събирането и обработването на достоверна информация (управление на информационни потоци), свързана с възможност за мониторинг на процесите при обучение на докторанти .

203

**КОНЦЕПТУАЛНО МОДЕЛИРАНЕ И ПРОЕКТИРАНЕ НА СИСТЕМАТА**

Изготвена е методика за разработването на софтуер за управление на информационни потоци, свързани с административното обслужване на докторанти във ВУЗ, съобразно законодателството на Република България. Въпреки, че тя е разработена за целите на настоящата статия, части от нея биха били приложими за изработването, на която и да е друга сходна система.

Първоначалните стъпки, които разработчика трябва да изпълни за правилното проектирането на един софтуер са: *Провеждане на интервю* с бъдещите потребители (клиенти), *Дефиниране на предназначението* на системата и *Дефиниране на задачите*, които системата трябва да изпълнява [3].

Задачите, които трябва да изпълнява настоящата информационна система са:

- Поддръжка на пълен набор от данни за един докторант;
- Възможност за отразяване на събитието, един докторант да има повече от един научен ръководител;
- Възможност за промяна на темата на дисертационния труд, като се пази информация и за предходните;
- Възможност за отразяване на учебната и научно-изследователската дейност за всяка учебна година;
- Възможност за генериране на справки и отчети за ръководните органи, както и такива изисквани от МОН.

Следва кратко описание на използваната методика за разработване на информационна система за управление на информационни потоци, свързани с административното обслужване на докторанти.

Започва се с етап „*Проучване*", в който се анализират и изучават съществуващите законови разпоредби както в ЗВО [2] така и в Правилника за образователната дейност в съответния университет [4, 5]. Разглеждат се съществуващи софтуерни продукти, удовлетворяващи частично законовите разпоредби.

В етап „*Спецификация*" се извършва формулиране на изискванията към сигурността на данните, към архитектурата на софтуера, към графичния потребителски интерфейс и към хардуера.

В етап „*Моделиране*" се извършва изготвяне на различните видове модели на софтуерния продукт.

В етап „*Планиране*" се уточняван избора на хардуерна и софтуерна платформа, на средата за разработване и на система за управление на бази от данни.

В етап „*Проектиране*" се извършва проектирането на релационната база от данни и нейната сигурност, и модулите на софтуерния продукт.

В етап „*Разработване*", се извършва разработването на отделните модули.

В етап „Интегриране" се извършва интегриране на отделните модули в цялостен софтуерен продукт и извършване на предварителното му тестване. Ако възникнат някакви грешки по време на тестването, се преминава в началото на етап „Проектиране", с цел корекция.

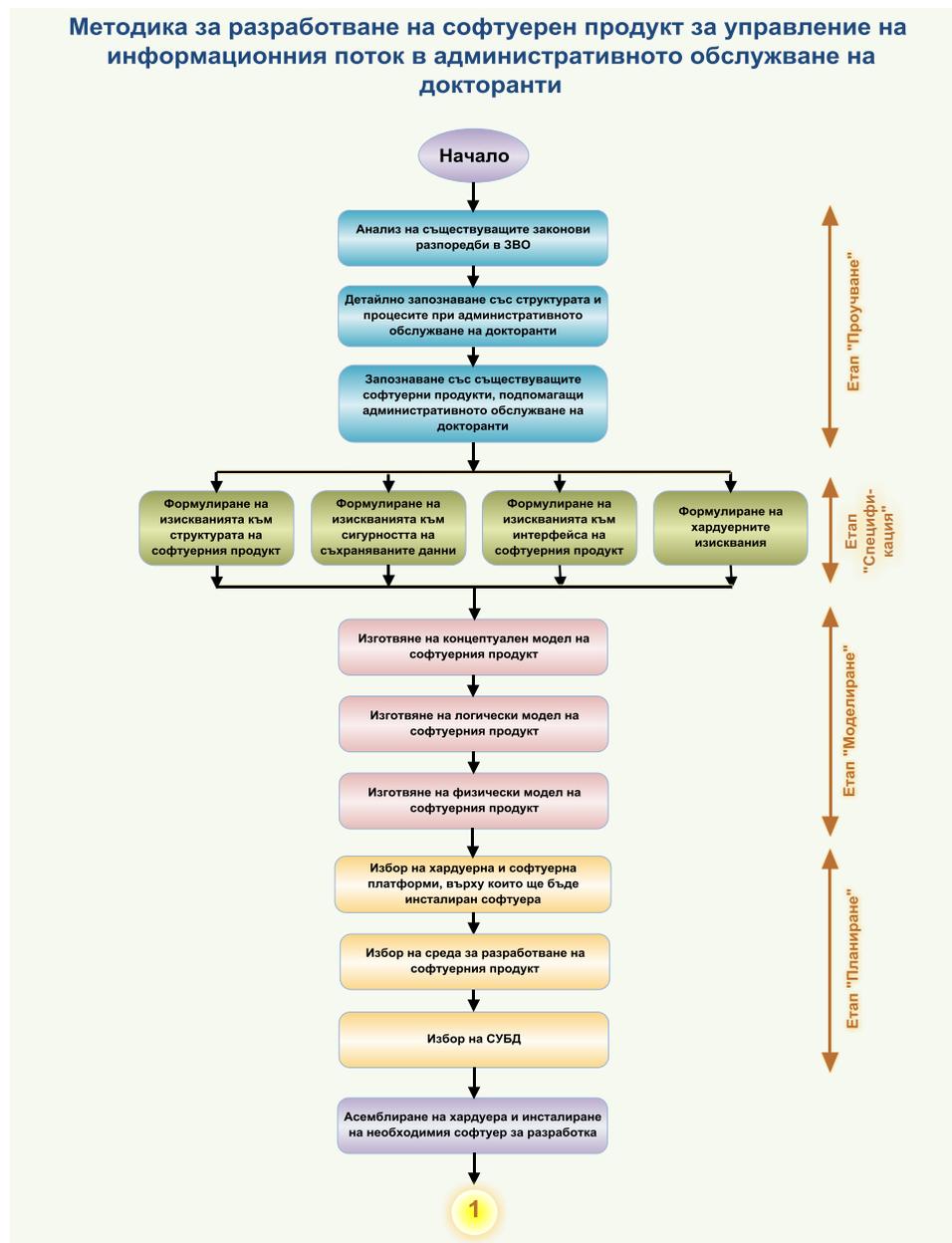

*Фигура 1*. Методика за изграждане на софтуер за управление на потоци от данни

205

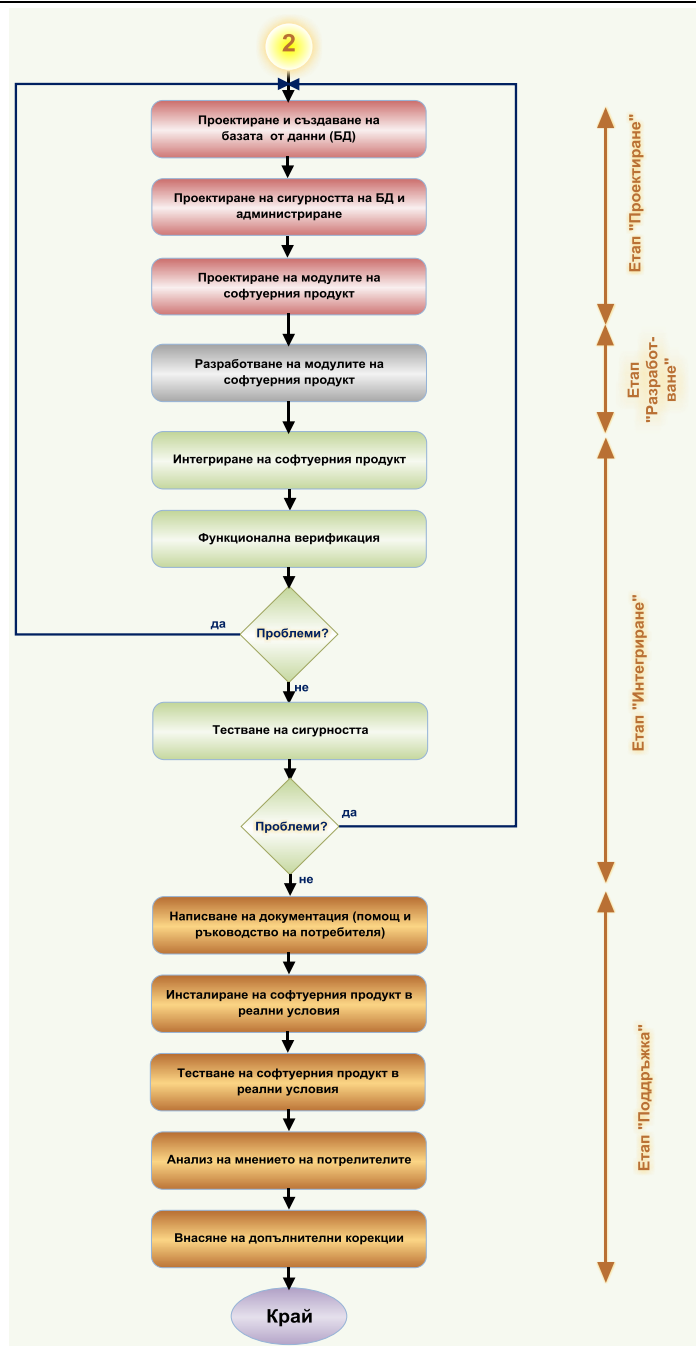

*Фигура 2*. Методика за изграждане на софтуер за управление на потоци от данни

206

Етап „Поддръжка", включва инсталиране на информационната система върху съответните клиентски машини; анализ на мнението на потребителите; изготвяне на крайната документация; и цялостно тестване.

Описаната методика е представена схематично на фигура 1 и фигура 2.

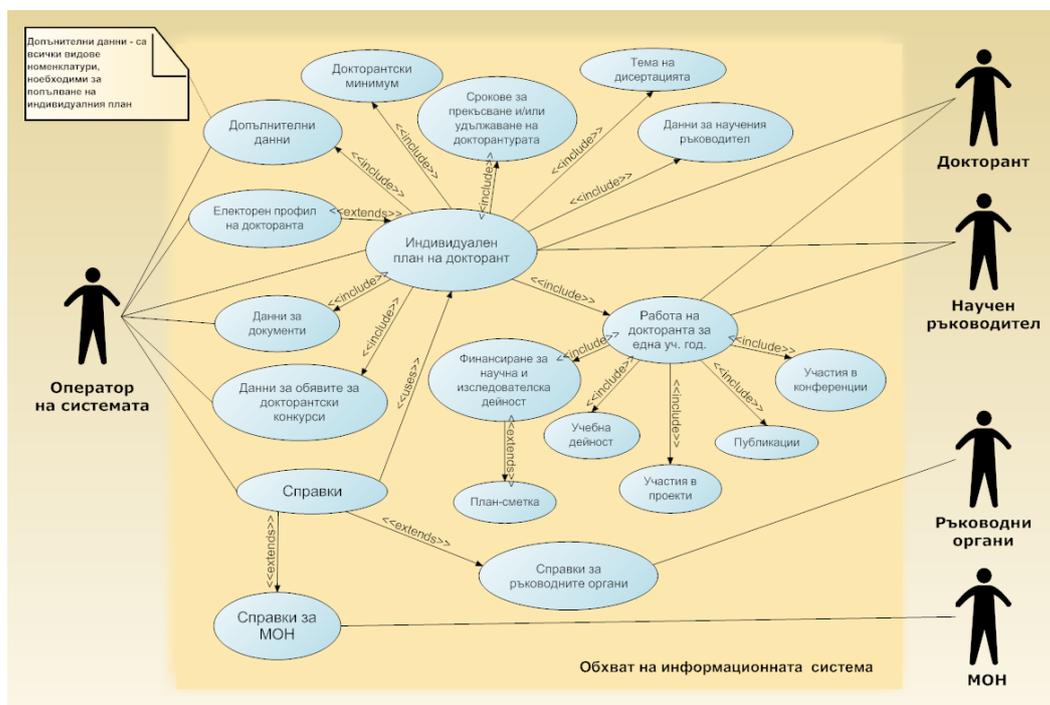

*Фигура 3.* Концептуален модел на архитектурата на софтуер за управление на информационни потоци

След направеното проучване на основните процеси свързани с приемането и обучението на докторантите съгласно Закона за висшето образования (ЗЗО) се обобщи, че информационната система е необходимо да има възможност за въвеждане и поддържане на следната информация:

- Данни за докторантите;
- Данни за хабилитираните лица, които са научни ръководители;
- Данни за необходимите документи;
- Данни за обявени докторантски конкурси;
- Справки за МОН;
- Индивидуален работен план на докторанта:
    o Вид на докторантурата;
    o Дата на зачисляване;
    o Срок на докторантурата;

207

- o Дата на отчисляване
- o Тема на дисертационен труд;
- o Научен ръководител;
- o Основание за приемане;
- o Данни за докторантски минимум;
- o Защита на дисертационен труд;
- o Научна дейност на докторанта по учебни години:
  - Участия в конференции;
  - Разработени проекти;
  - Проведени специализации;
  - Публикувани статии;
  - Участия в учебната дейност - проведени семинарни и лабораторни упражнения;

Концептуалния модел на информационна система за административно обслужване на докторанти във ВУЗ е представена на фигура 3.

**ЕКСПЕРИМЕНТАЛНО ИЗСЛЕДВАНЕ НА ЕФИКАСНОСТТА НА КОНЦЕПТУАЛНИЯ МОДЕЛ**

Въз основа на предложената методология беше разработен софтуер за административно обслужване на докторанти. За разработването на графичния потребителски интерфейс е използвана средата за визуално и събитийно-ориентирано програмиране Turbo C++ Builder® for Microsoft® Windows™ версия Explorer. На фигура 4 е представен изглед от графичния потребителски интерфейс.

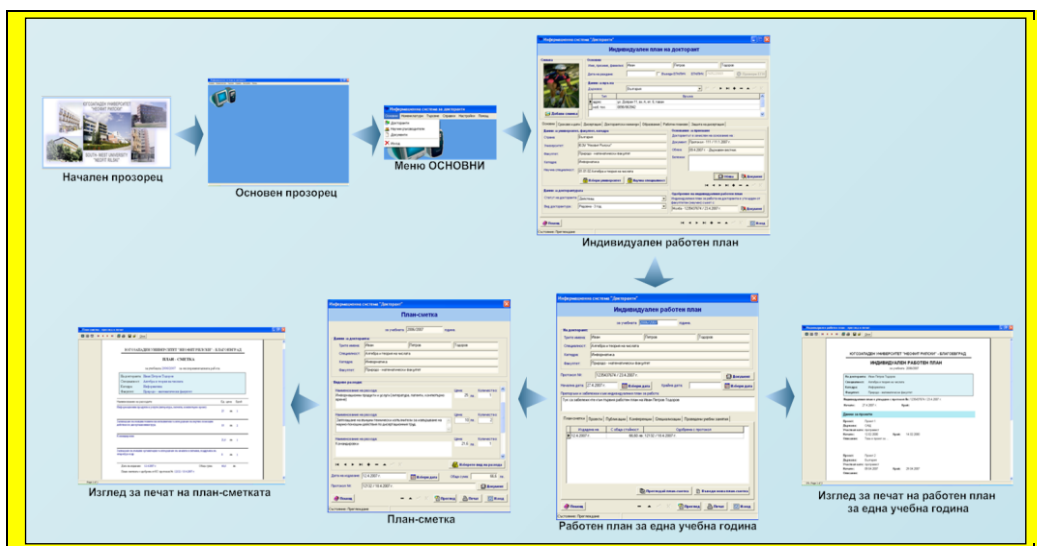

*Фигура 4*. Изглед на част от графичния потребителски интерфейс на информационната система.

208

Разгледаната информационна система е внедрена в ЮЗУ „Неофит Рилски" – гр. Благоевград, отдел „Конкурси". Системата до този момент се държи адекватно. Беше подложена на редица тестове, чрез които се провери нейната сигурност, и надеждността на обработваната информация.

**ЗАКЛЮЧЕНИЕ**

В настоящата статия бяха представени последователно процесите на анализиране, разработване на концептуален модел, методология на разработване и внедряване на реална информационна система, осигуряваща административното обслужване на докторанти във ВУЗ.

При направеното изследване са постигнатите следните резултати:

- Създадена е методика за проектиране на софтуер за управление на информационни потоци;
- Разработен е концептуален модел;
- Разработена е информационна система за административно обслужване на докторанти във ВУЗ;
- Същата е внедрена в ЮЗУ „Неофит Рилски", гр. Благоевград

**ЛИТЕРАТУРА**

**ЗА АВТОРИТЕ**

Радослава Кралева, Югозападен университет „Неофит Рилски", Природо-математически факултет, катедра „Информатика", ул. "Иван Михайлов" № 66, Благоевград 2700, e-mail: rady_kraleva@swu.bg

Велин Кралев, Югозападен университет „Неофит Рилски", Природо-математически факултет, катедра „Информатика", ул. "Иван Михайлов" № 66, Благоевград 2700, e-mail: velin_kralev@swu.bg

# ON MODEL OF INFORMATION SYSTEM FOR MANAGEMENT OF INFORMATION FLOWS

Radoslava Kraleva, Velin Kralev

***Abstract:*** *In this article are discussed some problems in developing software related to the management of information flows. We presented the basic stages in their development. We bold a methodology for conceptual modeling and design of information systems of this type. In order to demonstrate the effectiveness of the proposed model is an information system for administrative services of graduate students in the university.*

**Keywords:** information flow, software, conceptual model

## ABOUT AUTORS

Radoslava Kraleva, South-West University "Neofit Rilski", Faculty of Mathematics and Natural Sciences, Department of Informatics, 66 Ivan Michailov str.,2700 Blagoevgrad, Bulgaria, e-mail: rady_kraleva@swu.bg

Velin Kralev, South-West University "Neofit Rilski", Faculty of Mathematics and Natural Sciences, Department of Informatics, 66 Ivan Michailov str.,2700 Blagoevgrad, Bulgaria, e-mail: velin_kralev@swu.bg